\documentclass[MSNbibl,nameyear,dvips]{arxstspdf}
\usepackage{flushend}
\usepackage{stfloats}

\makeatletter

\def\@bmisc[#1]{%
  \get@battribute{unstr}%
  \common@pub@types%
  \let\bauthor\bbl@bauthor%
  \let\bhowpublished\@firstofone%
  \def\borganization##1{{\bauthor@style ##1}}%
}

\makeatother

\volume{26}
\issue{1}
\pubyear{2011}
\firstpage{12}
\lastpage{14}
\doi{10.1214/11-STS337A}
\referstodoi{10.1214/10-STS337}

\begin{document}
\begin{frontmatter}

\title{Discussion of ``Statistical Inference: The~Big Picture'' by R.~E.~Kass}
\runtitle{Discussion}
\pdftitle{Discussion of Statistical Inference: The~Big Picture by R.~E.~Kass}

\begin{aug}
\author{\fnms{Steven N.} \snm{Goodman}\corref{}\ead[label=e1]{sgoodman@jhmi.edu}}

\runauthor{S. N. Goodman}

\affiliation{Sidney Kimmel Comprehensive Cancer Center}

\address{Steven N. Goodman is Professor of Oncology, Epidemiology and Biostatistics,
Sidney Kimmel Comprehensive Cancer Center, Department of Oncology, Division of Biostatistics/Bioinformatics,
550 N. Broadway, Suite 1103, Baltimore, Maryland 21205, USA \printead{e1}.}

\end{aug}

% ABSTRACT

% KEYWORDS

\end{frontmatter}

In this piece, Rob Kass brings to bear his insights from a long career
in both theoretical and applied statistics to reflect on the disconnect
between what we teach and what we do. Not content to focus just on
didactic and professional matters, the focus of his 2009 article (Brown
and Kass, \citeyear{BroKas09}), in this commentary he proposes a remake of the
foundations of inference. He proposes to replace two fundamental
``isms''---frequentism and Bayesianism---with a new
``ism''---``pragmatism;'' an approach that he puts forward as more ecumenical
and practical, enshrining in foundations what good statisticians already
do.

There is a lot to commend in this piece, particularly the emphasis on
the subjunctive nature of all model-based inference, and I am sure the other
commentators will do justice to its strengths. But in spite of its
clarity and initial promise, I found Kass's proposal ultimately
unsatisfying. It seems less a new foundational philosophy than a call
for a truce, one of many over the years. It is telling that all of the
examples show practical equivalence between Bayesian and frequentist
estimates, so the biggest stakes here seem to be what people think, not
what they do. The difficulty with ``big tent'' foundations is that in
circumstances where different philosophies within the tent dictate
different actions, there is no guidance as to what route to take.

It is interesting to contrast this with the philosophic version of
``pragmatism,'' originally put forth by the polymath C. S. Peirce in the
late 1800s [also credited with proposing the log-likelihood ratio as a~%
measure of evidence (Hacking, \citeyear{Hac65})], whose intellectual heirs included
William James, Thomas Dewey, W. V. O. Quine and Richard Rorty. Pragmatism
embraced three maxims, the most important of which was that the meaning
of ideas was defined by their practical, observable effects. Ideas that
made no material difference in the real world had no meaning.

Kass alludes to a possible difference in real-world consequences just
once, in his mention of the analysis of high-dimensional data sets. But
he states his pragmatist philosophy is agnostic on how to approach
these, and that the choice should be ``according to their performance
under theoretical conditions thought to capture relevant real-world
variation in a particular applied setting.'' It would have been
extraordinarily useful to see such an example, and if indeed there could
be a model-based resolution of what are often quite difficult
conundrums.

In the domain with which I am most familiar, clinical trials, the
traditional frequentist-Bayesian inferential dilemma arises most
commonly in the interpretation of ``early stopped'' trials, that is to
say, should the inference depend upon the stopping rule, and if so, how?
This becomes particularly acute when the stopping is due to an unplanned
analysis. This particular situation arose recently in the high-profile
case of the diabetes drug Avandia. In 2007, a meta-analysis was
published that raised concern about the cardiovascular risks of Avandia
(Nissen and Wolski, \citeyear{NisWol07}), leading to calls that the FDA should remove
the drug from the market. The RE\-CORD trial was being conducted in Europe
to examine the efficacy and safety of Avandia, and its industry sponsor
requested an unplanned analysis in response to the new data. This
analysis (arguably) indicated no excess cardiac risk, and this interim
result was then published, at the behest of the sponsor (Home et al.,
\citeyear{Hometal07}; Nissen, \citeyear{Nis10}). Many doubted that an interim result that had
demonstrated excess risk would have been published and discounted the
result. How should this be sorted out? What are the dimensions of
``real-world variation'' here that we should include in the model, and
on what grounds do we determine how to measure the evidence and how to
act? This was a real decision, with real, big-time consequences. What
guidance would ``statistical pragmatism'' give us in skirmishes like
that?

Kass says he has been guided by ``past and present sages,'' but leaves
the job of naming them to Barnett (\citeyear{Bar99}). Kass is right that many have
preceded him on this path, and it would have been quite illuminating to
compare this proposal more directly to those of his predecessors. Among
the several I would have liked to have seen contrasted would be George
Box, whose descriptions of the theoretical nature and pragmatic utility
of models, which he adjoined to his attempts to resolve the
Bayesian-frequentist ``deadlock,'' are remarkably similar to this essay
in spirit, if not in substance. Box's paper on Fisher's ecumenicism,
``Science and Statistics'' (Box, 1976), included graphics not so
different than those found here, albeit with a more prominent role for
experimental ``filter'' through which we see the world. He thought it
important that our ``wrong'' models be subject to continual revision in
accord with changes in scientific understanding, one of Kass's central
points.

Interestingly, in a response to a 1990 essay by Glen Shafer in this
journal on the ``Unity and Diversity of Probability,'' which had similar
aims to this one (Shafer, \citeyear{Sha90}), Box (\citeyear{Box90}) stated ``There is another
substantive issue I would like to raise. This concerns the fatal
fascination of the word `unity.' Unity in many things is desirable, but
we should not be trying to impose `oneness' on a situation where
`twoness' is of the essence.'' One wonders if he would make
the same comment here. He went on to characterize statistical inference
as reducible to model fitting and model criticism, claiming that
Bayesians are better at fitting, frequentists better at criticism, and
that we need to be good at both.

It is interesting to look to the field of bioethics to see how it deals
with a variety of foundational theories seemingly at odds with each
other, but which capture important features of desirable ethical
conduct. Two dominant theories are utilitarian (or consequentialist),
which focus on outcomes (somewhat akin to frequentist approaches) and
deontologic,\break which focus on the intrinsic morality of how people treat
each other, which has some parallels to Bayesian logic (Beauchamp and
Childress, \citeyear{BeaChi01}). These underlying ethical philosophies provided, as all
philosophies must, competing definitions for foundational concepts,
which in ethics include moral\break goods and moral duties.

However, in looking for ethical principles that\break should guide clinical
research, ethicists borrowed from multiple traditions, enshrining that
guidance in the Belmont report, which did not try to resolve
foundational differences (National Commission, \citeyear{autokey8}).

The three principles it espoused were (1) Respect for persons
(autonomy), a primarily deontologic construct; (2) Beneficence (and
non-maleficence), mainly utilitarian, and (3) Justice (fair distribution
of harms and benefits), derived from yet other moral philosophies. These
principles are sometimes in tension, reflecting their different
foundational pedigrees, keeping ethicists in business. But the Belmont
principles have still proven to be enormously influential and useful,
capturing the key features of each theory, translating them into the
applied domain, and providing a framework for regulation and for ethical
debate.

So perhaps Kass's proposal might be best framed not as a statistical
philosophy, but as the beginning of a code governing statistical conduct
and teaching. It could embrace such things as a desire for Bayesianly
coherent procedures that have good frequentist properties, and perhaps
provide guidance on the kinds of difficult questions that Kass has posed
in his previous writings (Kass, \citeyear{Kas06}). Such a code need not resolve
foundational differences, as statistical pragmatism does not, but it can
distill the desiderata of those philosophies down to a kernel of
principles applicable to all applied problems, implicitly endorsing
goals of competing philosophies that most would support. I can see all
the pieces of such a code here, but it would take some further work to
abstract them.

In summary, I welcome this as an insightful piece with admirable goals,
most of which I and, I suspect, other statisticians share. But
whether those goals are best met by replacing our foundational theories,
or by distilling and collectively endorsing the aspects of those
theories that are most useful, is an open question. I believe that there
are difficult inferential and decision problems that defy any
foundational attempt at resolution, and that their origins are found
outside statistics, in the incompleteness of our substantive knowledge.
This creates the gap Kass highlights between our theoretical models and
reality, a gap that certainly deserves to be front and center in any
conversations about statistical procedures or results. But we still must
draw conclusions and take action. I did not find an improved guide to
such actions in this piece, but I did appreciate its renewed call to not
let foundational dogma determine which direction we take. If this piece
can serve as a step toward doing for statistics what the Belmont Report
did for research ethics, it will have served a~very important role.

\end{document}